\documentclass[aps,twocolumn,showpacs]{revtex4}
\usepackage{epsf}
\def\beq{\begin{equation}}
\def\eeq{\end{equation}}

\def\bk{{\bf k}}
\def\bk'{{\bf k}'}

\begin{document}
%
\title{Rescattering and finite formation time effects  in 
electro-disintegration of the deuteron in the cumulative region}
\author{M. A. Braun}
\affiliation{ Department  of High Energy Physics, S. Petersburg University, 
198904 S. Petersburg,
 Russia}
\author{C. Ciofi degli Atti}
 \author{L.P. Kaptari}
 \altaffiliation{On leave from  Bogoliubov Lab. Theor. Phys.,141980, JINR,  Dubna, Russia}
  \affiliation{Department of Physics, University  of Perugia and INFN Sezione di Perugia,
       via A. Pascoli, Perugia, I-06123, Italy}
%
\begin{abstract}
 The role of rescattering due to the final state interaction (FSI)  and the 
influence of the finite formation time (FFT)
  on
 the inclusive $D(e,e')X$ and exclusive $D(e,e'p)n$ electro-disintegration of
 the deuteron 
are studied  in the cumulative
kinematical region $x>1$ and moderate values of the 4-momentum transfer  
$Q^2=2\div 10$ (GeV/c)$^2$.
The spins are averaged out. It is found that in the inclusive process the 
relative magnitude
of rescattering steadily grows with $x$ and that at  $x=1.7$ it has the same
order as the plane wave impulse approximation (PWIA) contribution, with 
the 
finite formation time
effects decreasing  the rescattering contribution by $\sim 30$\%.  In the 
exclusive process, with increasing momentum transfer,
 FFT   substantially reduces the effects from FSI, although the latter 
 are still appreciable in the region of momentum transfer investigated.
\end{abstract}
%
\maketitle
\section{Introduction}
\label{intro}
High-energy electro-disintegration of the deuteron
is a powerful tool to investigate, 
first,  how and when at larger energies and momentum 
transfers the
description in terms of hadrons (nucleons and mesons) transforms into
the one in terms of quarks and gluons  and, second,
the Colour Transparency (CT) effects predicted by 
QCD.
In terms of relevant Feynman 
diagrams,  one expects that whereas at comparatively low energies 
 virtual nucleons and mesons with standard propagators can be used, at large
enough virtualities such a description  gradually becomes invalid.
To clearly see this phenomenon one has to be able to reach high enough
virtualities in the process, which can be achieved 
by choosing the kinematics forbidden  for free (on-mass-shell)
nucleons, i.e. the so-called {\it cumulative} kinematics, which
corresponds to values of the Bjorken scaling variable larger than one (
note that the importance of studying the $x$-dependence of CT was first
stressed in Ref. ~\cite{kop}).
In the PWIA the cross-sections
are directly related to the behaviour of the deuteron  wave
function.  However such a direct
relation is broken by  FSI, which naively are expected to be  large
in the cumulative region; it is here that CT effects are to be taken
into account. CT predicts that at high $Q^2$  FSI become small,
so that, in principle, choosing both $x$ and $Q^2$ large enough one
can neglect FSI and have access, via the PWIA,  to the high-momentum behaviour of the
nuclear wave function. It is of particular importance that for high values 
 of  $x$ and $Q^2$  the rescattering energy may be quite small
~\cite{ciofi}, so that  only two-nucleon intermediate 
states can be considered and the simplest  Feynman diagrams
with only nucleon lines (although at high virtuality) have to be evaluated.
In such a kinematical situation CT effects can originate from the virtual excitation of the ejectile.

In this study we investigate the relative role of FSI and  CT effects
 in the deuteron electro-disintegration, 
using an approximate picture in which
the spins are  altogether averaged out. Taking 
the relativistic spin into account,
apart from purely technical difficulties, inevitably introduces
a variety of unknown off-shell components of both the
electromagnetic production vertex and the rescattering
amplitude, which deprives the results of any predictive power.
To relate our absolute cross-sections to the experimental data
we    introduce an 
effective electromagnetic form factor, chosen to reproduce the
PWIA results with full relativistic spins ~\cite{ciofi}. In our approach
 CT is introduced on the hadronic level using its
  equivalence to the final formation time (FFT) effect
~\cite{BCT}.

\section{Basic formulas}

Separating the leptonic part, the process we are going to study is
\beq
\gamma^*(q)+d(2p)\to N(p_1)+N(p_2),
\label{process1}
\eeq
where 4-momenta are denoted in brackets.
The total c.m. energy squared in the process is 
\begin{equation}
s=(2p+q)^2=Q^2\frac{2-x}{x}+M^2,\ \ x=\frac{Q^2}{2qp},
\label{esse}
\end{equation}
where $M$ is the deuteron mass.
The cumulative region we are going to study corresponds to
$
1<x<2.
$
From Eq. (\ref{esse}) one concludes that at $x$ close to 2
the c.m. energy remains close to the threshold even if $Q^2$ is large, 
so that only the lowest
two-nucleon intermediate states can be considered. The calculation of the cross-section for the
process (\ref{process1}) then reduces to the evaluation of the square modulus of 
the two standard diagrams shown
in Fig. 1, which correspond to the PWIA and rescattering contributions, respectively.
\begin{figure}[h]
\epsfxsize 3.5in
\centerline{\epsfbox{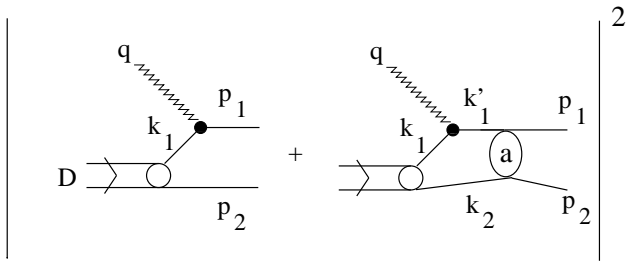}}
\caption{ The square of the sum of
the PWIA and  rescattering amplitudes 
proportional to the  cross sections.}
\label{Fig1}
\end{figure}

Our final expression for the inclusive process
\beq
e(l)+d(2p)\to e(l')+X
\label{process2}
\eeq
can be standardly expressed through the hadronic tensor $W_{\mu\nu}$,
which is just the square of the two amplitudes shown in Fig.1
integrated over the momenta of the two final nucleons. It can be shown that
in order to obtain the exclusive
cross-section for the lepto-disintegration,
averaged over the azimuthal angles, it is sufficient to drop this 
integration.

The PWIA contribution to the
amplitude is just the 
hadronic matrix element
of the electromagnetic current
\beq
J_{\mu}^{PWIA}=2\gamma(Q^2)K_{\mu}G(k_1^2),
\label{incl}
\eeq
where   $\gamma$ is the effective electromagnetic form-factor
of the (scalar) nucleon, the 
vector $K$ is chosen to guarantee conservation of the electromagnetic current, and 
$k_1=2p-p_2=p_1-q$ is the momentum of the active 
nucleon before the interaction.
The  quantity  $G(k_1^2)$ is the relativistic deuteron 
wave function, which in principle
can  be sought as a solution of the 
Bethe-Salpeter equation  for the
deuteron. However  the relativistic potential for this equation is
unknown at high momentum transfers and virtualities, 
so rather than use forms for this potential  fitted to comparatively
low energy and momentum transfer scattering, we directly approximate 
$G$ using as a guide its non-relativistic limit.  
To calculate the rescattering contribution
we choose a system with $q_{\perp}=0$ (``lab'' system for 
reaction  (\ref{process1}))
which simplifies the integrations over the intermediate
nucleon momenta.  In this system the integration over $k_{2-}$
puts either the spectator or the active nucleon on the mass-shell, depending on 
whether $k_{2+}$ is lower  or higher  than $P_++q_+$. In both cases,
as in  the PWIA contribution,
the integrand involves only the relativistic deuteron wave function
with one of the nucleons  on its mass shell. 

 As a result we find  the rescattering contribution to the matrix
element of the hadronic current in the form
\beq
J_{\mu}^{resc}=2\gamma(Q^2)\int 
dV(k_1,k_2)K_{\mu}G(v)\frac{a(k_2|p_2)}{m^2-(k_1+q)^2},
\label{jresc}
\eeq
where $k_1+k_2=2p$, $K=k_1+q(k_1q)/Q^2$ and  $a$ is the rescattering amplitude, with 
$dV$ and $v$ having  different forms 
in the two mentioned regions of integrations. 
All invariant arguments entering the integrand in Eq. 
(\ref{jresc}) have to be expressed through the light-cone integration
variables, taking into account that either $k_2^2=m^2$, or
$k_1^2=m^2$.

The total hadronic tensor is obtained as
\beq
W_{\mu\nu}=\int dV(p_1,p_2)(J_\mu^{PWIA}+J_\mu^{resc})
(J_\nu^{PWIA}+J_\nu^{resc})^*,
\label{wtot}
\eeq
where $dV(p_1,p_2)$ is the standard invariant phase volume for the
produced nucleons. 
As far as CT is concerned, we introduce it via the FFT of the hit hadron, which manifests itself through a  dependence
of the scattering amplitudes and verteces on the virtuality of the hit hadron after $\gamma*$ absorption (see [3]).
We assume that the  effect of this dependence
 can be 
modelled by a monopole form-factor,
which is equivalent to changing the ejectile propagator as follows
\beq
\frac{1}{m^2-k^{\prime\,2}_1}\to\frac{1}{m^2-k^{\prime\,2}_1}-
\frac{1}{{m^*}^2-k^{\prime\,2}_1}.
\label{denom}
\eeq 
This substitution is equivalent to
assuming that there are two different ejectile states with
masses $m$ and $m^*$ whose contribution to rescattering cancels
out in the limit of high momenta, in agreement with the underlying ideas of  CT [3].
Following  Refs. ~\cite{BCT},~\cite{FGMS},  we choose  for the mass $m^*$
the value 1.8 GeV.

\section{Numerical calculations}

\begin{figure}[h]
\epsfxsize 3.0in
\vskip 3mm
\centerline{\epsfbox{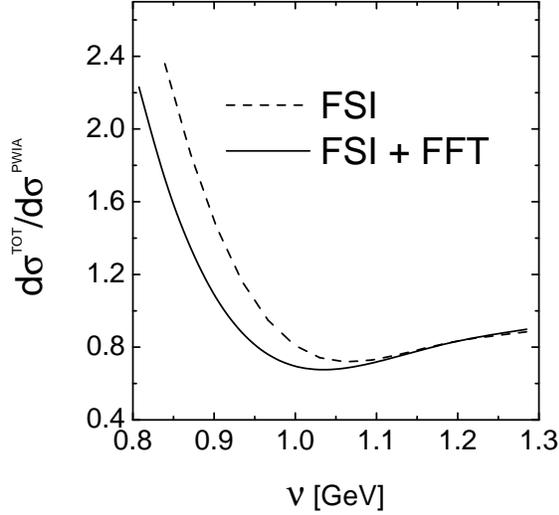}}
\caption{ Ratio of  the total inclusive cross section $d\sigma^{TOT}\equiv 
d\sigma^{TOT}/dE'd\Omega'$ to the PWIA cross section 
$d\sigma^{PWIA}\equiv 
d\sigma^{PWIA}/dE'd\Omega'$.
 In the dashed curve  $d\sigma^{TOT}$ includes only rescattering
 effects, whereas in the full curve  both rescattering and FFT effects
 are present.}
\label{Fig2}
\end{figure}
\vskip 5mm

We parametrize the relativistic deuteron wave function $G(k^2)$ using as a guide
the  nucleon density in the deuteron at comparatively low momenta. In the
non-relativistic limit one obtains 
\beq
G^2(k^2)=2M(2\pi)^3|\Psi({\bf k}^2)|, \ \ 
{\bf k}^2=\frac{1}{2}(m^2-k^2-M\epsilon),
\eeq
where $\epsilon$ is the deuteron binding energy. For $|\Psi|^2$ we have taken
the form which corresponds to the AV14 interaction  ~\cite{pot}.
As for the rescattering amplitude, it was chosen  in the form
\beq
a(s,t)=(\alpha+i)\sigma^{tot}(s)\sqrt{s(s-4m^2)}e^{bt},
\eeq
with the values of the parameters  $\sigma^{tot}$, $\alpha$ and $b$
taken from ~\cite{ampl}.

As already mentioned, our main goal was to estimate the magnitude of the FSI
and the influence of CT (or, equivalently, FFT) in the cumulative region.
Accordingly, our basic quantities to be calculated  are the  ratios of the
rescattering to the  PWIA contributions. We expect that the
error due to  neglecting  spins is reduced in these ratios.
Still, to have a clearer physical picture and to be able to compare with the 
experimental data, we also tried to calculate  absolute values of the 
cross-sections. To this end we introduced an effective electromagnetic
form-factor for the active nucleon, chosen to approximately take into 
account  magnetic interaction of both the proton and neutron.
Comparison of
our impulse approximation 
with the results obtained with a full account of  spins [2]
leads to the choice
\beq
\gamma^2(Q^2)=\gamma_D^2(Q^2)\frac{2+\tau (\mu_p^2+\mu_n^2)}{1+\tau},
\label{formfactor}
\eeq
where $\gamma_D$ is the standard
dipole form-factor,
$\tau=Q^2/(4m^2)$, and $\mu_{p,n}$ are the anomalous magnetic moments of
the proton and neutron. With this choice our PWIA results practically coincide 
with the ones with full relativistic spins taken into account. 

The inclusive cross-sections for the process (\ref{process2}) 
have been calculated at points corresponding to  the
experimental  data of ~\cite{exp},  with the initial electron energy
$E=9.761$ GeV and  scattering  angle $\theta=10^o$. We have considered
values for the final electron energy which cover the region of $x$
in the interval $1.0<x<1.71$. Some relevant kinematical values 
characterizing   the
chosen points are listed in Table 1. 

\begin{figure}[h]
\epsfxsize 3.5in
\vskip -10mm
\centerline{\epsfbox{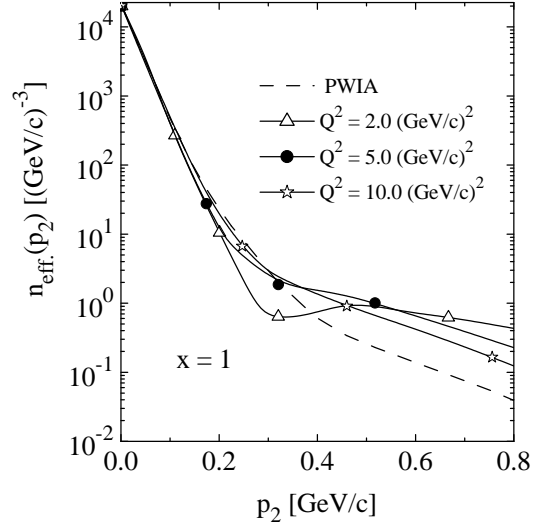}}
\vskip -1.5in
\caption{ The effective momentum distribution
(Eq. \ref{ratio1})  {\it vs.}
 the neutron recoil momentum 
 ${|\bf p}_2| \equiv p_2$ at $x=1$ . The dot-dashed curve represents the PWIA result, 
 whereas the other curves include also rescattering and FFT effects at various values of  $Q^2$.}
\label{Fig3}
\end{figure}
\begin{table}[h,t]
\begin{center}
\caption{Some kinematical variables in the inclusive electro-disintegration
cross section corresponding to the kinematics of of ~\cite{exp}, {\it viz}
incident electron energy $E=9.761 GeV$ and scattering angle $\theta = 10^o$.
$\nu$ is the energy transfer, $x$ the Bjorken scaling variable, $Q^2$ the square 
4-momentum transfer, $s$ the produced invariant mass, and $p_{lab}$ the 
the momentum of the struck nucleon in the system where the
spectator is at rest $p_{lab}$. Note that the inelastic  threshold 
corresponds to $s \simeq 4 \,\,GeV^2$ ($p_{lab}\simeq \,\,0.8\,\, GeV$) (cf. Ref. \cite{ciofi}).}

\vskip 2mm

\begin{tabular}{|c|c|c|c|c|}\hline
           &        &      & & \\
$\nu$, GeV &    x   & $Q^2,\, (GeV/c)^2$ &$s,\, GeV^2$ &$p_{lab},\,GeV/c$ \\
           &        &      &    &     \\\hline
0.826      &1.71    &2.65  &3.96   & 0.73    \\
0.872      &1.61    &2.64  &4.14   & 0.88    \\
0.930      &1.50    &2.62  &4.38   & 1.06    \\
0.987      &1.41    &2.60  &4.61   & 1.21    \\
1.056      &1.30    &2.58  &4.89   & 1.40    \\
1.137      &1.20    &2.56  &5.22   & 1.61   \\
1.228      &1.10    &2.53  &5.58   & 1.82   \\
1.332      &1.00    &2.50  &6.00   & 2.07   \\\hline
\end{tabular}
\end{center}
\vskip 2mm
\label{table1}
\end{table}
\noindent It can be seen, 
that at high
values of $x$ (high cumulativity) $p_{lab}$ is small, well below
the threshold energy
for pion production. 
This justifies our approach based on the assumption that only nucleon 
degrees of freedom
are relevant; however
at smaller $x$ $p_{lab}$ grows and such an assumption becomes of
disputable validity.

\begin{figure}[h]
\epsfxsize 2.5in
\centerline{\hspace*{-10mm}\epsfbox{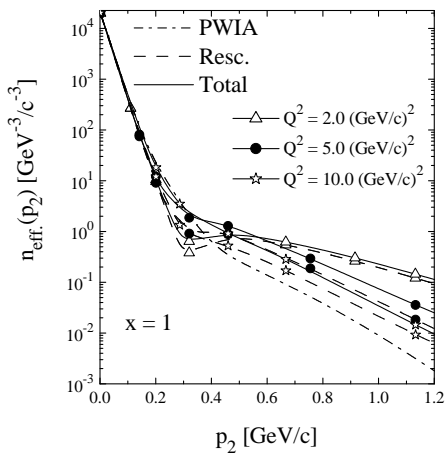}}
\caption{ The effective momentum distribution
(Eq. (\ref{ratio1}))  at $x=1$  which includes
rescattering and FFT effects (full),  and rescattering effects only (dashed); the difference
between  the full and dashed curves is due to FFT effects; the dot-dashed curve represents the
PWIA result.}
\label{Fig4}
\end{figure}

In Fig. 2 the ratios of the full cross section (which includes FSI
with or without FFT)  to  the PWIA cross section is shown.
It can be seen 
that the rescattering contribution, which is  very small at $x\sim 1$,
steadily grows with  $x$, reaching an
order of about 50\% already at $x\sim 1.3$. The relative role of FFT (or CT)
also rises with $x$. At the maximum value of $x$ studied, $x=1.71$, FFT effects
decrease  FSI by $\sim 30$\%. This however does not make  FSI
smaller than the PWIA contribution, so that they cannot be neglected at all.

 
Let us now discuss  the exclusive cross-section $d(e,e')pn$.
In order to minimize  the error of neglecting spins, we consider the reduced
cross-section
\beq
n_{eff}(|{\bf p}_2|) \equiv n_{eff}(|{\bf p}_2|,x,Q^2)=
|\Psi(|{\bf p}_2|)|^2\frac{\sigma_{excl}}{\sigma_{excl}^{PWIA}},
\label{ratio1}
\eeq
\noindent
 which  in  PWIA  reduces to   the input nucleon momentum distribution
in the deuteron. Here $p_2$ is the momentum of the unobserved nucleon
(the missing momentum).
Fig. 3   shows the effective momentum distributions {\it vs} the missing
momentum at $x=1$, calculated taking rescattering and FFT into account.
Calculations have been performed  
at fixed values of  $x$ and $Q^2$ and  different  missing momenta $\bf p_2$.
Due to  energy conservation, this corresponds to different values of
the angle  between  $|{\bf q}|$ and  $|{\bf p}_2|$.

\begin{figure}[h]
\epsfxsize 2.7in
\vskip -1mm
\centerline{\hspace*{-10mm}\epsfbox{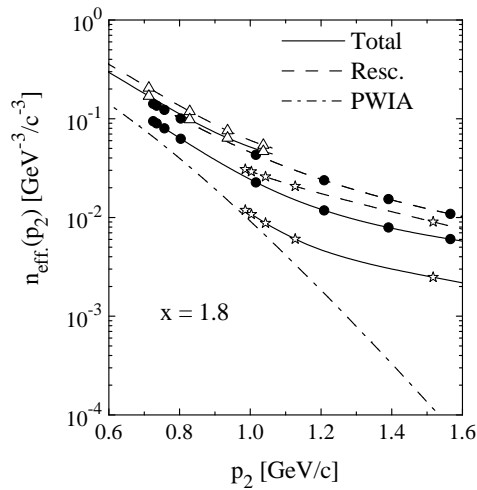}}
\caption{ The same as in Fig. \ref{Fig4} at $x=1.8$}
\label{Fig5}
\end{figure} 
In Figs. 4 and 5 the effects of rescattering and FFT 
are shown separately at $x=1$,  and in the deep kinematical region, $x=1.8$. 
It can be seen
from Fig.4 that at $x=1$ 
FFT effects, as expected,  
increase with $Q^2$,
in agreement with the results obtained for the process $^4He(e,e'p)^3H$
(~\cite{mor}).
At $x=1.8$
 (Fig. 5) one observes that
 with the growth of the missing momentum,  FSI  appreciably 
decrease due to FFT, which  makes
the distorted cross-section more similar to the PWIA one.

\section{Conclusions}

Using the Feynman diagram technique we have calculated
the rescattering contribution to the  cross-section of  inclusive
and exclusive deuteron electro-disintegration, paying
particular attention to the cumulative region  $x>1$.
Our main goal has been to study
 CT effects, which we have  introduced via
the FFT of the struck nucleon. In our calculations spins has been
averaged out and all particles were treated as scalar ones. The error 
introduced by this approximation is expected to cancel to a large extent 
in  the ratio of the total cross-section (which includes FSI and FFT effects),  with and without FSI
and FFT effects, to the PWIA cross section. To be able to compare absolute magnitudes of the cross-sections
with the experimental data we used an effective electromagnetic vertex for 
the active nucleon, chosen to describe  the PWIA results with spins.

The results of our calculations show that the relative magnitude of the 
FSI steadily grows with $x$. At $Q^2\sim 2$ (GeV/c)$^2$ and  $x\sim 1.7$,
the rescattering 
contribution raises the cross-section nearly by a factor of 2.5, whereas the  introduction
of FFT decreases back the cross section  by a factor of  1.6, which is evidently not enough
to disregard FSI altogether. As expected, at large values of $x$ the influence of FFT grows up:
it can be seen  from Fig. 5  that at $Q^2=10$ (GeV/c)$^2$ 
and $|{\bf p}_2|\sim 1 \div 1.5$ GeV/c,  FFT decreases the pure FSI 
results by a factor of 4.

In conclusion, we have found that the CT or FFT effects are clearly
visible in the electro-disintegration of the deuteron in the
cumulative region, which may serve as a tool for their experimental study.
We have also found that although FFT effects decrease the effects of the  FSI rather substantially
at $Q^2=  2 \div 10 (GeV/c)^2$,
this is by far not sufficient to neglect FSI altogether.
\section{Acknowledgments}
 
 This work was partially supported by the Ministero dell'I\-stru\-zio\-ne, Universit\`{a} e Ricerca (MIUR), 
through the funds COFIN01.
M. B. and L.P.K. are  indebted to  the University of
Perugia and INFN, Sezione di Perugia, for warm hospitality and financial support.




\begin{thebibliography}{100}
\bibitem{kop} B.K. Jennings and B.Z. Kopeliovich, Phys. Rev. Lett.,
{\bf 70} (1993) 3384
%
\bibitem{ciofi} 
C. Ciofi degli Atti, L.P. Kaptari and D. Treleani, 
Phys.Rev. {\bf C63} (2001) 044601
%
\bibitem{BCT}
 M. A. Braun, C. Ciofi degli Atti and D. Treleani, Phys. Rev.
{\bf C62} (2000) 034606
%
\bibitem{FGMS}
 L.L. Frankfurt, W.R. Greenberg, G.A. Miller and M.I. Strikman,
Phys. Rev., {\bf C46} (1992) 2547
%
\bibitem{pot} R.B. Wiringa, R. A. Smith and T. L. Ainsworth, Phys. Rev.
{\bf C29} (1984) 1207
%
\bibitem{ampl} R. A. Arndt et al., "Partial-Wave Analysis Facility
(SAID)", http:/said.phys.vt.edu
%
\bibitem{exp} S. Rock {\it et al}, Phys. Rev. Lett.{\bf 49} (1982) 1139;\\
R. G. Arnold {\it et al}, {\it ibid.}
{\bf 61}(1988) 806
%
\bibitem{mor} H. Morita, M. A. Braun, C. Ciofi degli Atti, and D.
Treleani, Nucl. Phys. {\bf A699} 
(2002) 328c
\end{thebibliography}
\end{document}